\documentstyle[12pt]{article}
\title{Making the small oblique parameters large}
\author{L.\ Lavoura\thanks{On leave of absence
from Universidade T\'ecnica de Lisboa,
Lisbon, Portugal}\hspace{1mm}
and Ling-Fong Li \\
\small Department of Physics, Carnegie-Mellon University, \\
\small Pittsburgh, Pennsylvania 15213, U.S.A.}
\begin{document}
\maketitle
\begin{abstract}
We compute the oblique parameters,
including the three new parameters $ V $,
$ W $ and $ X $  introduced recently by the Montreal group,
for the case of one scalar multiplet
of arbitrary weak isospin $ J $ and weak hypercharge $ Y $.
We show that,
when the masses of the heaviest and lightest
components of the multiplet remain constant,
but $ J $ increases,
the oblique parameter $ U $ and the three new oblique parameters
increase like $ J^3 $,
while $ T $ only increases like $ J $.
For large multiplets with masses not much higher than $ m_Z $,
the oblique parameters $ U $ and $ V $ may become much larger
than $ T $ and $ S $.
\end{abstract}

\vspace{5mm}

The oblique parameters characterize the influence
of physics beyond the standard model on the experimentally
measurable quantities in terms of their  contributions
to the usual gauge-boson propagators.
The first oblique parameter considered by the theorists
\cite{T} was $ T $,
which in many cases is larger than the other oblique parameters,
because of its quadratic,
instead of logarithmic,
dependence on the masses of the new particles.
Later \cite{S},
the parameter $ S $ was introduced,
and it was shown that,
because $ S $ receives a significant mass-independent contribution
from each chiral fermion doublet,
the experimentally allowed range for that parameter
severely constrains technicolor theories.
A more general parametrization of the effects of new physics,
under the assumption that the mass scale of that physics
is much higher than the $ Z $ mass,
which implies that the vacuum-polarization functions
are approximately linear in $ p^2 $ up to $ p^2 = m_Z^2 $,
requires one more oblique parameter,
$ U $ \cite{STU}.
Recently \cite{beyondSTU},
the above assumption has been relaxed,
and it has been shown that,
whatever the mass scale of the new physics may be,
the fact that all precision electroweak measurements
are made at either $ p^2 = 0 $ or $ p^2 = m_Z^2 $
(or,
in the case of $ m_W $ and $ \Gamma_W $,
at $ p^2 = m_W^2 $),
requires the use of a total of six oblique parameters,
thereby introducing three new parameters,
which were named $ V $,
$ W $ and $ X $.
The definitions of the six oblique parameters
given in reference \cite{beyondSTU} are
\begin{eqnarray}
S & \equiv &
\frac{16 \pi c_w^2 s_w^2}{e^2}\,
\left[
\frac{A_{ZZ} (M_Z) - A_{ZZ} (0)}{M_Z}
- \frac{c_w^2 - s_w^2}{c_w s_w}
\left. \frac{\partial A_{\gamma Z} (P)}{\partial P} \right|_{P=0}
-
\left. \frac{\partial A_{\gamma \gamma} (P)}{\partial P} \right|_{P=0}
\right]\, ,
\label{eq:S}\\
T & \equiv &
\frac{4 \pi}{e^2}\,
\left[
\frac{A_{WW} (0)}{M_W} - \frac{A_{ZZ} (0)}{M_Z}
\right]\, ,
\label{eq:T}\\
U & \equiv &
\frac{16 \pi s_w^2}{e^2}\,
\left[
\frac{A_{WW} (M_W) - A_{WW} (0)}{M_W} \right.
\nonumber\\
  &   &
\left.
- c_w^2 \frac{A_{ZZ} (M_Z) - A_{ZZ} (0)}{M_Z}
- 2 c_w s_w
\left. \frac{\partial A_{\gamma Z} (P)}{\partial P} \right|_{P=0}
- s_w^2
\left. \frac{\partial A_{\gamma \gamma} (P)}{\partial P} \right|_{P=0}
\right]\, ,
\label{eq:U}\\
V & \equiv &
\frac{4 \pi}{e^2}\,
\left[
\left. \frac{\partial A_{ZZ} (P)}{\partial P} \right|_{P=M_Z}
- \frac{A_{ZZ} (M_Z) - A_{ZZ} (0)}{M_Z}
\right]\, ,
\label{eq:V}\\
W & \equiv &
\frac{4 \pi}{e^2}\,
\left[
\left. \frac{\partial A_{WW} (P)}{\partial P} \right|_{P=M_W}
- \frac{A_{WW} (M_W) - A_{WW} (0)}{M_W}
\right]\, ,
\label{eq:W}\\
X & \equiv &
\frac{4 \pi c_w s_w}{e^2}\,
\left[
\left. \frac{\partial A_{\gamma Z} (P)}{\partial P} \right|_{P=0}
- \frac{A_{\gamma Z} (M_Z)}{M_Z}
\right]\, .
\label{eq:X}
\end{eqnarray}
Here as in the rest of this paper,
we use capital letters to denote squared masses:
$ M_Z \equiv m_Z^2 $,
and so on.
Also,
$ P \equiv p^2 $.
The letter $ A $ refers to the coefficient of $ g_{\mu \nu} $
in the vacuum polarization.
Note that the new parameters $ V $,
$ W $ and $ X $ vanish when the vacuum polarizations
are linear functions of $ P $.
They therefore include the effects \cite{grinstein}
of second and higher derivatives relative to $ P $,
or equivalently of terms in the functions $ A(P) $
which contain masses of new particles M in the denominator
($ P^2 / M $,
$ P^3 / M^2 $ and so on).
Thus these three parameters must decrease
when the masses of the new particles increase.
It was shown in references \cite{beyondSTU} and \cite{fit} that,
without any approximations,
the oblique contributions to all the quantities which may be measured
with high precision at either zero-momentum transfer,
or at the $ Z $ pole,
may be written in terms of the four parameters $ S $,
$ T $,
$ V $ and $ X $.
The contributions to $ m_W $ and to $ \Gamma_W $
also involve the parameters $ U $ and $ W $.
Furthermore,
in reference \cite{fit} a fit to all these quantities
using the available data has been performed,
yielding that the precision for five of the oblique parameters
is quite good (with error bars of order $ \pm 1 $),
while the parameter $ W $ is only poorly known,
requiring a precise measurement of $ \Gamma_W $.

It is generally believed that $ S $ and $ T $ will be larger than all
other oblique parameters.
$ T $ varies quadratically with the masses of the new particles,
and an approximate custodial symmetry is needed to prevent $ T $
from becoming excessively large.
Once the approximate custodial symmetry is present,
$ U $ also becomes small.
Indeed,
in the limit of exact custodial symmetry,
$ U $,
just as $ V $,
$ W $ and $ X $,
only receives contributions from the fact that the functions $ A (p^2) $
are not linear in $ p^2 $
(and from the gauge coupling of hypercharge \cite{T},
{\it i.e.},
from the fact that $ c_w^2 \neq 1 $ and $ M_W \neq M_Z $).
On the other hand,
$ S $ receives,
in the fermion case,
a contribution which,
being independent of the fermion masses,
does not vanish in the limit of custodial symmetry.
Therefore,
we expect $ S $ to be dominant
in the case of exact custodial symmetry,
as displayed in a particular example in Figure 2
of reference \cite{beyondSTU};
and we expect $ T $ to become comparable or larger than $ S $ when
there is a large breaking of custodial symmetry.

The purpose of this paper is to point out
that $ S $ and $ T $ are not always larger than the other parameters
as the previous picture suggests,
and there are cases where the other parameters can be much larger.
We have computed the oblique parameters for the case of
an arbitrary scalar multiplet,
with weak isospin $ J $ and weak hypercharge $ Y $.
We find that,
in general,
for sufficiently large $ J $
and for sufficiently low masses of the components of the scalar multiplet,
$ T $ and $ S $ may remain only moderately large,
while the parameters $ U $ and $ V $ increase rapidly
with the size of the multiplet (that is,
with $ J $),
becoming the largest oblique parameters in certain situations.
We do not know of any particular theoretical motivation for
introducing large scalar multiplets with masses of order $ m_Z $.
However, it is useful to keep in mind that {\it a priori} all the
oblique parameters can be large,
and therefore not to assume the smallness of those parameters
when attempting fits of the experimental data.

We consider a complex scalar multiplet with weak isospin $ J $.
For simplicity,
we assume that this multiplet does not develop
vacuum expectation value (VEV),
and that the $ 2 J + 1 $ components
of this multiplet do not mix among themselves
(which might happen if two components of the multiplet
had the same electric charge)
nor with any other scalars in the theory.
For the moment,
we also assume that these components
all have different and arbitrary masses.
We denote the third component of isospin
of each component by $ I $,
and its squared mass by $ M_I $.
The photon couples to each of the scalars with strength $ e (I + Y) $,
while the $ Z $ couples with strength
$ e (I c_w^2 - Y s_w^2) / (c_w s_w) $.
The $ W $ couples the field with third component of isospin $ I $
with the field with third component of isospin $ (I - 1) $,
the strength of the coupling being
$ e \sqrt{(J + I)(J - I + 1)} / (\sqrt{2} s_w) $,
as is well known from the theory of angular momentum.
Therefore,
the one-loop vacuum-polarization functions are given by
\begin{eqnarray}
A_{\gamma \gamma} (P) & = &
\frac{e^2}{8 \pi^2} \sum_{I=-J}^{+J} (I+Y)^2 F(M_I, M_I, P)\, ,
\label{eq:gammagamma}\\
A_{\gamma Z} (P) & = &
\frac{e^2}{8 \pi^2 c_w s_w} \sum_{I=-J}^{+J} (I+Y) (I c_w^2 - Y s_w^2)
F(M_I, M_I, P)\, ,
\label{eq:gammaZ}\\
A_{ZZ} (P) & = &
\frac{e^2}{8 \pi^2 c_w^2 s_w^2} \sum_{I=-J}^{+J} (I c_w^2 - Y s_w^2)^2
F(M_I, M_I, P)\, ,
\label{eq:ZZ}\\
A_{WW} (P) & = &
\frac{e^2}{16 \pi^2 s_w^2} \sum_{I=-J}^{+J} (J^2 - I^2 + J + I)
F(M_I, M_{I-1}, P)\, .
\label{eq:WW}
\end{eqnarray}
Notice that,
just as $ A_{\gamma \gamma} (0) $ and $ A_{\gamma Z} (0) $
vanish because of electromagnetic gauge invariance,
in this particular case $ A_{ZZ} (0) $ also vanishes \cite{li}.
This happens because,
as shown below,
$ F(M_I, M_I, 0) = 0 $.
The function $ F $ may be computed from the relevant Feynman diagrams,
which are depicted in Figure 1.
One obtains
\begin{eqnarray}
F(M_1, M_2, P) & = & {\rm div}.P + \frac{M_1 \ln y_1 + M_2 \ln y_2}{2}
\nonumber\\
               &   &
- \int_0^1 dx [P x^2 + (M_1 - M_2 - P) x + M_2]
\nonumber\\
               &   &
\times \ln [y_P x^2 + (y_1 - y_2 - y_P) x + y_2]\, .
\label{eq:functionF}
\end{eqnarray}
Here,
div is a divergent quantity typical of dimensional regularization,
the exact definition of which is irrelevant,
because it will cancel in the combinations for the oblique parameters.
Also,
$ y_1 = M_1 / \mu^2 $,
$ y_2 = M_2 / \mu^2 $ and $ y_P = P / \mu^2 $,
where $ \mu $ is the arbitrary mass parameter
used in dimensional regularization,
which will also cancel in the final results.
We need the following properties of the function $ F $ in order
to calculate the oblique parameters:
\begin{equation}
4 F(M_1, M_2, 0) = \theta_+ (M_1, M_2) \equiv
M_1 + M_2 - \frac{2 M_1 M_2}{M_1 - M_2} \ln \frac{M_1}{M_2}\, ;
\label{eq:Fzero}
\end {equation}
\begin{equation}
\left. \frac{\partial F(M_1, M_1, P)}{\partial P} \right|_{P=0} =
{\rm div} + \frac{1}{6} + \frac{1}{6} \ln \frac{M_1}{\mu^2}\, ;
\label{eq:Fderivative}
\end{equation}
\begin{equation}
{\rm div} + \frac{1}{6} + \frac{1}{12} \ln \frac{M_1 M_2}{\mu^4}
- \frac{F(M_1, M_2, M) - F(M_1, M_2, 0)}{M} =
\xi (\frac{M_1}{M}, \frac{M_2}{M})\, ;
\label{eq:Fcsi}
\end{equation}
\begin{equation}
\left. \frac{\partial F(M_1, M_2, P)}{\partial P} \right|_{P=M}
- \frac{F(M_1, M_2, M) - F(M_1, M_2, 0)}{M} =
\rho (\frac{M_1}{M}, \frac{M_2}{M})\, .
\label{eq:Frho}
\end{equation}
The functions $ \xi (x, y) $ and $ \rho (x, y) $ vanish when
both $ x $ and $ y $ tend to zero.
The function $ \xi $ is defined by
\begin{eqnarray}
\xi (x, y) & \equiv &
\frac{4}{9} - \frac{5}{12} (x + y) + \frac{1}{6} (x-y)^2
\nonumber\\
            &  &
+ \frac{1}{4} \left[ x^2 - y^2 - \frac{1}{3} (x-y)^3 -
\frac{x^2 + y^2}{x - y} \right] \ln \frac{x}{y}
\nonumber\\
            &  &
- \frac{1}{12} \Delta(x, y) f(x, y)\, ,
\label{eq:csi}
\end{eqnarray}
where
\begin{equation}
\Delta(x, y) \equiv - 1 + 2 (x + y) - (x - y)^2\, ,
\label{eq:Delta}
\end{equation}
and
\begin{equation}
f (x, y) \equiv \left\{ \begin{array}{ll}
- 2 \sqrt{\Delta (x, y)} \left(
\arctan \frac{x - y + 1}{\sqrt{\Delta (x, y)}}
- \arctan \frac{x - y - 1}{\sqrt{\Delta (x, y)}} \right)
& \mbox{if $ \Delta (x, y) > 0 $} \\
0
& \mbox{if $ \Delta (x, y) = 0 $}\, . \\
\sqrt{- \Delta (x, y)} \ln
\frac{x + y - 1 + \sqrt{- \Delta (x, y)}}{x + y - 1 - \sqrt{- \Delta (x, y)}}
& \mbox{if $ \Delta (x, y) < 0 $}
\end{array} \right.
\label{eq:f}
\end{equation}
The function $ \rho $ is defined by
\begin{eqnarray}
\rho (x, y) & \equiv &
\frac{1}{6} - \frac{3}{4} (x + y) + \frac{1}{2} (x-y)^2
\nonumber\\
            &  &
+ \frac{1}{4} \left[ 2 (x^2 - y^2) - (x-y)^3
- \frac{x^2 + y^2}{x - y} \right] \ln \frac{x}{y}
\nonumber\\
            &  &
- \frac{1}{4} [x + y - (x - y)^2] f(x, y)\, .
\label{eq:rho}
\end{eqnarray}
The expressions for $ \xi (x, x) $ and for $ \rho (x, x) $
are obtained by performing the limit $ y \rightarrow x $
in Eqs.~\ref{eq:csi} and \ref{eq:rho}.
Also,
$ \theta_+ (M, M) = 0 $.

It is straightforward to get the following expressions for
the new oblique parameters $ V $,
$ W $ and $ X $:
\begin{eqnarray}
V & = & \frac{1}{2 \pi c_w^2 s_w^2} \sum_{I=-J}^{+J}
(I c_w^2 - Y s_w^2)^2 \rho (\frac{M_I}{M_Z}, \frac{M_I}{M_Z})\, ,
\label{eq:Vresult}\\
W & = & \frac{1}{4 \pi s_w^2} \sum_{I=-J}^{+J}
(J^2 - I^2 + J + I) \rho (\frac{M_I}{M_W}, \frac{M_{I-1}}{M_W})\, ,
\label{eq:Wresult}\\
X & = & \frac{1}{2 \pi} \sum_{I=-J}^{+J}
(I + Y) (I c_w^2 - Y s_w^2) \xi (\frac{M_I}{M_Z}, \frac{M_I}{M_Z})\, .
\label{eq:Xresult}
\end{eqnarray}

For convenience,
we separate the contributions to each of the parameters
$ S $ and $ U $ into two pieces:
$ S = S^{\prime} + S^{\prime \prime} $ and
$ U = U^{\prime} + U^{\prime \prime} $.
The second terms characterize the deviation
from linear dependence in $ P $ of $ A_{ZZ} (P) $ and $ A_{WW} (P) $:
\begin{eqnarray}
S^{\prime \prime} & = & - \frac{2}{\pi} \sum_{I=-J}^{+J}
(I c_w^2 - Y s_w^2)^2 \xi(\frac{M_I}{M_Z}, \frac{M_I}{M_Z})\, ,
\label{eq:S''}\\
U^{\prime \prime} & = & \frac{1}{\pi} \sum_{I=-J}^{+J}
\left[
2 (I c_w^2 - Y s_w^2)^2 \xi(\frac{M_I}{M_Z}, \frac{M_I}{M_Z})
- (J^2 - I^2 + J + I) \xi(\frac{M_I}{M_W}, \frac{M_{I-1}}{M_W})
\right]\, .
\label{eq:U''}
\end{eqnarray}
The other contributions to $ S $ and $ U $ are
\begin{eqnarray}
S^{\prime} & = & - \frac{Y}{3 \pi} \sum_{I=-J}^{+J} I \ln \frac{M_I}{\mu^2}\, ,
\label{eq:S'}\\
U^{\prime} & = & \frac{1}{6 \pi} \sum_{I=-J}^{+J} (J^2 + J - 3 I^2)
\ln \frac{M_I}{\mu^2}\, .
\label{eq:U'}
\end{eqnarray}
Notice that,
because of the relations
\begin{equation}
\sum_{I=-J}^{+J} I = \sum_{I=-J}^{+J} (J^2 + J - 3 I^2) = 0\, ,
\label{eq:nullsum}
\end{equation}
the dependence of $ S^{\prime} $ and $ U^{\prime} $ on $ \mu^2 $
cancels out,
as expected.
Finally,
for $ T $ we have the result
\begin{equation}
T = \frac{1}{16 \pi c_w^2 s_w^2 M_Z}
\sum_{I=-J}^{+J} (J^2 - I^2 + J + I)\, \theta_+ (M_I, M_{I-1})\, ,
\label{eq:Tresult}
\end{equation}
in which $ M_W = c_w^2 M_Z $ has been used.

We first check that the above results agree with
the decoupling theorem \cite{decoupling}.
There is an SU(2)$\otimes$U(1)-invariant mass term
which gives a common squared-mass $ M $ to all the members of the multiplet.
The mass splittings of the multiplet,
on the other hand,
come from quartic couplings to the usual  Higgs doublet
which gets a VEV,
$ v $.
We therefore write $ M_I = M + \delta_I $.
The decoupling theorem \cite{decoupling} tells us that,
when $ M $ goes to infinity,
with the $ \delta_I $ remaining constant
(because they are proportional to $ v^2 $,
which is held fixed),
all the oblique parameters must vanish.
In fact,
we find in this limit,
\begin{eqnarray}
\theta_+ (M_I, M_K) & = &
\frac{1}{3} \frac{(\delta_I - \delta_K)^2}{M}
- \frac{1}{6} \frac{(\delta_I - \delta_K)^2 (\delta_I + \delta_K)}{M^2}
+ O(\frac{1}{M^3})\, ,
\label{eq:decoupletheta}\\
\xi (\frac{M_I}{M_X}, \frac{M_K}{M_X}) & = &
\frac{1}{60} \frac{M_X}{M}
+ \left[ \frac{M_X^2}{840} - \frac{M_X (\delta_I + \delta_K)}{120}
- \frac{(\delta_I - \delta_K)^2}{60} \right] \frac{1}{M^2}
+ O(\frac{1}{M^3})\, ,
\label{eq:decouplexi}\\
\rho (\frac{M_I}{M_X}, \frac{M_K}{M_X}) & = &
- \frac{1}{60} \frac{M_X}{M}
+ \left[ - \frac{M_X^2}{420} + \frac{M_X (\delta_I + \delta_K)}{120}
\right] \frac{1}{M^2}
+ O(\frac{1}{M^3})\, .
\label{eq:decouplerho}
\end{eqnarray}
This shows that $ T $,
$ V $,
$ W $,
$ X $,
$ S^{\prime \prime} $ and $ U^{\prime \prime} $
all go to zero when $ M $ increases.
As for $ S^{\prime} $ and $ U^{\prime} $,
they vanish when all the scalars of the multiplet have the same mass,
because of Eq.~\ref{eq:nullsum},
and therefore they must also be $ \sim M^{-1} $.
This establishes decoupling.
Eq.~\ref{eq:decoupletheta} was first noted in reference \cite{li}.

We now study explicitly the mass spectrum of the multiplet with
$ 2J + 1$ components.
We assume the existence of only one Higgs doublet,
$ H $,
and denote the additional scalar multiplet by $ \Delta $.
For a general hypercharge of $ \Delta $,
there are only two quartic couplings of $ \Delta $ with $ H $.
The first one may be written $ (\tilde{\Delta} \Delta)_1
(\tilde{H} H)_1 $.
Here,
$ \tilde{H} $ is the doublet conjugate to $ H $,
and $ \tilde{\Delta} $ is the multiplet conjugate to $ \Delta $.
In the quartic coupling written above,
the doublet $ \tilde{H} $ couples with the doublet $ H $
to form an SU(2)$\otimes$U(1) singlet,
which then multiplies the singlet formed from $ \Delta $
together with $ \tilde{\Delta} $.
This coupling just gives a common mass to all the components of
$ \Delta $,
when $ H $ develops a VEV.
The other quartic coupling is
$ (\tilde{\Delta} \Delta)_3 (\tilde{H} H)_3 $.
Here,
$ \tilde{\Delta} $ and $ \Delta $
pair into a triplet of SU(2),
and $ \tilde{H} $ and $ H $ do likewise.
Then the two triplets couple to form a gauge singlet.
Now,
\begin{equation}
(\tilde{\Delta} \Delta)_3 (\tilde{H} H)_3 =
\sum_{k=1}^3 (\Delta^{\dagger} T_k^{(J)} \Delta)
(H^{\dagger} T_k^{(1/2)} H)\, ,
\label{eq:coupling}
\end{equation}
in which the $ T_k^{(J)} $ are the generators of SU(2)
in the spin-$J$ representation.
Since it is the combination $ (H^{\dagger} T_3^{(1/2)} H) $ which gets a VEV,
we see that this quartic coupling will give
a contribution to the squared mass of each component
proportional to its $I$-value.
That is,
\begin{equation}
M_I = M + I \delta = \frac{M_2 + M_1}{2} + I \frac{M_2 - M_1}{2 J}\, ,
\label{eq:masses}
\end{equation}
in which we have denoted $ M_J $ by $ M_2 $,
and $ M_{-J} $ by $ M_1 $.
{}From now on,
we will assume the masses of the components of $ \Delta $
to be given by Eq.~\ref{eq:masses}.
This equation is very useful,
because it reduces the {\it a priori} $ 2 J + 1 $ independent masses
to only two independent ones.

Let us now consider what happens in the case
where we keep $ M_1 $ and $ M_2 $ fixed,
and let $ J $ increase.
That is,
we consider different scalar multiplets,
all with equally spaced squared masses,
and all with the same maximum and minimum mass,
but with a growing number $ 2 J + 1 $ of components,
and therefore with the spacing between the squared masses decreasing.
Obviously,
provided both $ M_1 $ and $ M_2 $ are higher than $ M_Z / 4 $,
this possibility is not excluded by experiment.
One asks,
what will happen to the oblique parameters when $ J $ increases?
First consider $ S^{\prime} $ and $ U^{\prime} $.
{}From Eq.~\ref{eq:masses},
we can write
\begin{equation}
\ln \frac{M_I}{\mu^2} = \ln \frac{M_2 + M_1}{2 \mu^2} +
\ln \left( 1 + \frac{I}{J} \frac{M_2 - M_1}{M_2 + M_1} \right)\, ,
\label{eq:logarithm}
\end{equation}
and the first term in the right-hand-side
contributes neither to $ S^{\prime} $ nor to $ U^{\prime} $,
due to Eq.~\ref{eq:nullsum}.
If we develop the last logarithm in
Eq.~\ref{eq:logarithm} as a power series in $ (M_2 - M_1) $,
and then perform the summation in Eq.~\ref{eq:S'},
we find
\begin{equation}
S^{\prime} = - \frac{Y}{3 \pi} \sum_{n=0}^{\infty}
\frac{1}{2 n + 1}
\frac{1}{J^{2 n + 1}}
\left( \frac{M_2 - M_1}{M_2 + M_1} \right)^{2 n + 1}
\sum_{I=-J}^{+J} I^{2 n + 2}\, .
\label{eq:S'series}
\end{equation}
Similarly,
we get for $ U^{\prime}$
\begin{equation}
U^{\prime} = - \frac{1}{6 \pi} \sum_{n=1}^{\infty} \frac{1}{2n}
\frac{1}{J^{2 n}}
\left( \frac{M_2 - M_1}{M_2 + M_1} \right)^{2 n}
\left[ (J^2 + J) \sum_{I=-J}^{+J} I^{2 n}
- 3 \sum_{I=-J}^{+J} I^{2 n + 2} \right]\, .
\label{eq:U'series}
\end{equation}
Because $ \sum_{I=-J}^{+J} I^n $ is a polynomial in $ J $
of degree $ n+1 $,
these equations tell us that,
in the limit of large $ J $,
$ S^{\prime} $ grows like $ J^2 $,
while $ U^{\prime} $ grows like $ J^3 $.

Let us now find out how does $ T $ behave in the same limit.
We expand $ \theta_+ (M_I, M_{I-1}) $ in powers of $ (M_2 - M_1) $,
obtaining
\begin{eqnarray}
\theta_+ (M_I, M_{I-1}) & = &
(M_2 + M_1) \sum_{n=2}^{\infty}
\left( \frac{-1}{J} \right)^n \left( \frac{M_2 - M_1}{M_2 + M_1} \right)^n
\nonumber\\
                        & \times &
\left\{ \frac{1}{n+1}\, [ (I-1)^{n+1} - I^{n+1} ]
      + \frac{1}{n}\, (2 I - 1)\, [ I^n - (I-1)^n ] \right.
\nonumber\\
                        & &
\left. + \frac{1}{n-1}\, (I^2 - I)\, [ (I-1)^{n-1} - I^{n-1} ] \right\}\, .
\label{eq:Texpansion1}
\end{eqnarray}
The terms in curly brackets in the right-hand-side of the above equation
appear at first sight to be a polynomial in $ I $ of degree $ n $.
It is easy to check,
however,
that the coefficients of $ I^n $ and of $ I^{n-1} $ both vanish.
After some simplification,
we indeed find
\begin{equation}
\theta_+ (M_I, M_{I-1}) =
(M_2 + M_1) \sum_{n=2}^{\infty}\,
\frac{1}{J^n} \left( \frac{M_2 - M_1}{M_2 + M_1} \right)^n
\sum_{p=0}^{n-2}\, ( - I )^p\, \frac{(n-2)! (n-p-1)}{p! (n-p+1)!}\, .
\label{eq:Texpansion2}
\end{equation}
Note that the sum over $ p $ in the
right-hand-side of Eq.~\ref{eq:Texpansion2}
is symmetric under $ I \rightarrow 1-I $ for even $ n $,
but changes sign under $ I \rightarrow 1-I $ for odd $ n $.
Therefore,
upon the summation over $ I $ specified in Eq.~\ref{eq:Tresult},
the terms with odd $ n $ disappear,
and we obtain
\begin{equation}
T =
\frac{1}{16 \pi c_w^2 s_w^2} \frac{M_2 + M_1}{M_Z}
\sum_{n=1}^{\infty} t_{2n}
\left( \frac{M_2 - M_1}{M_2 + M_1} \right)^{2n}\, .
\label{eq:Tseries}
\end{equation}
The coefficients $ t_{2n} $ are the functions of $ J $
which are found by performing the relevant summations.
The important point is that,
in Eqs.~\ref{eq:Tresult} and \ref{eq:Texpansion2},
the maximum power of $ I $ to be summed is $ I^n $,
while there is a $ J^n $ in the denominator.
Therefore,
the coefficients $ t_{2n} $,
and $ T $,
increase linearly with $ J $ for large $ J $.

The implications of these facts can be seen in Figure 2,
where we plot $ T $,
$ S^{\prime} $ and $ U^{\prime} $ as functions of $ m_2 $,
while keeping $ m_1 $ fixed at 150 GeV.
The hypercharge $ Y $ of the multiplet is $ -1/2 $,
while $ J = 1, 3 $ and 5 in Figures 2a, 2b and 2c,
respectively.
Evidently,
in all three figures $ T = S^{\prime} = U^{\prime} = 0 $ when $ M_2 = M_1 $.
Also,
for large $ M_2 $,
$ T $ is much larger than both $ S^{\prime} $ and $ U^{\prime} $,
due to the dependence in $ (M_2 + M_1) / M_Z $
made explicit in Eq.~\ref{eq:Tseries}.
However,
for small $ M_2 $,
we see that,
for large $ J $,
$ U^{\prime} $ becomes much larger than $ T $.
This is because $ U^{\prime} $ increases like $ J^3 $,
while $ T $ increases only like $ J $.
$ S^{\prime} $ also becomes much larger than $ T $,
because $ S^{\prime} $ increases like $ J^2 $;
but it must be noted that $ S^{\prime} $ can anyway always be
either suppressed or enhanced by the hypercharge $ Y $,
because $ S^{\prime} $ is proportional to $ Y $.

Now consider the additional contributions $ S^{\prime \prime} $
and $ U^{\prime \prime} $ to $ S $ and $ U $,
respectively,
as well as the new oblique parameters $ V $,
$ W $ and $ X $.
All these parameters do not vanish when $ M_2 = M_1 $,
in contrast to what happens with $ T $,
$ S^{\prime} $ and $ U^{\prime} $.
Let us then calculate the oblique parameters when $ M_2 = M_1 = M $.
In this case,
$ T = 0 $ while
\begin{eqnarray}
\frac{S}{2 J + 1} & = & - \frac{2}{\pi}
\left[ \frac{J (J+1)}{3} c_w^4 + Y^2 s_w^4 \right]
\xi (\frac{M}{M_Z}, \frac{M}{M_Z})\, ,
\label{eq:Ssum}\\
\frac{U}{2 J + 1} & = & \frac{2}{\pi}
\left\{
\left[ \frac{J (J+1)}{3} c_w^4 + Y^2 s_w^4 \right]
\xi (\frac{M}{M_Z}, \frac{M}{M_Z})
- \frac{J (J+1)}{3}
\xi (\frac{M}{M_W}, \frac{M}{M_W})
\right\}\, ,
\label{eq:Usum}\\
\frac{V}{2 J + 1} & = & \frac{1}{2 \pi c_w^2 s_w^2}
\left[ \frac{J (J+1)}{3} c_w^4 + Y^2 s_w^4 \right]
\rho (\frac{M}{M_Z}, \frac{M}{M_Z})\, ,
\label{eq:Vsum}\\
\frac{W}{2 J + 1} & = & \frac{1}{2 \pi s_w^2}
\frac{J (J+1)}{3} \rho (\frac{M}{M_W}, \frac{M}{M_W})\, ,
\label{eq:Wsum}\\
\frac{X}{2 J + 1} & = & \frac{1}{2 \pi}
\left[ \frac{J (J+1)}{3} c_w^2 - Y^2 s_w^2 \right]
\xi (\frac{M}{M_Z}, \frac{M}{M_Z})\, .
\label{eq:Xsum}
\end{eqnarray}
Notice that $ U $ would vanish if $ c_w = 1 $ and $ M_W = M_Z $.
This is because the gauge coupling of hypercharge
violates custodial symmetry \cite{T}.
It is seen that all oblique parameters --- except $ T $ ---
increase like $ J^3 $ or,
more explicitly,
like $ J (J+1) (2J+1) $.
However,
the contributions proportional to $ Y^2 $ increase only like $ (2 J + 1) $.
It is easy to check that,
in the situation in which $ M_2 \neq M_1 $,
contributions to the oblique parameters
proportional to $ Y (M_2 - M_1) $ arise,
which increase like $ J^2 $.

This behavior is easy to understand.
In general,
we should expect a factor of $ J $ always
to be present in the oblique parameters,
just corresponding to the number of independent graphs in
the vacuum polarizations;
see Eqs.~\ref{eq:gammagamma} to \ref{eq:WW}.
Moreover,
as the coupling of a gauge boson of SU(2) to a scalar with third component
of isospin equal to $ I $ increases like $ I $,
and as there are two such vertices in a one-loop contribution
to a vacuum polarization,
we expect a further factor of $ J^2 $ in every oblique parameter,
corresponding to the square of the maximum value of $ I $ in the multiplet.
However,
for the coupling of the gauge boson of U(1),
we get a factor $ Y $ and lose a factor $ I $ (or,
in the final result after the summation over $ I $,
of $ J $).

Thus,
the behavior of $ T $,
which increases like $ J $ instead of like $ J^3 $,
is an exception to the rule.
This is  connected to the fact that $ T $ is quadratic,
instead of logarithmic,
in the masses of the particles.
For two particles with squared masses $ M $ and $ M^{\prime} $,
and small mass difference,
$ T $ will vary as $ (M - M^{\prime})^2 / (M + M^{\prime}) $,
while all other oblique parameters will vary as $ \ln (M / M^{\prime}) $.
Now,
in our procedure of letting the maximum and minimum squared masses
of the multiplet fixed while increasing the size of the multiplet,
a factor $ 1 / J $ is automatically associated
with each power of the mass difference $ M - M^{\prime} $.
This is what makes $ T $ increase with $ J $ instead of
increasing with $ J^3 $,
as all other oblique parameters.

In Figure 3 we present a plot of all six oblique parameters
as functions of $ m_2 $,
with $ m_1 = 150 $ GeV kept fixed,
for a fixed value of the hypercharge
and for three different values of $ J $.
It is seen that,
for large $ J $ and low $ m_2 $,
$ V $ becomes quite large.

\vspace{5mm}

This work was supported by the United States Department of Energy,
under the contract DE-FG02-91ER-40682.

\vspace{5mm}

%
%

\vspace{10mm}

%
\hspace{5mm} {\large \bf FIGURE CAPTIONS}

\vspace{10mm}

Figure 1: Feynman diagrams relevant for the computation of
the vacuum polarizations.

\vspace{5mm}

Figure 2: Plot of $ S^{\prime} $ (full line),
$ T $ (dotted line) and $ U^{\prime} $ (dashed line) against $ m_2 $,
with $ m_1 = 150 $ GeV kept fixed.
All figures are for a scalar multiplet with hypercharge $ -1/2 $.
Figure 2a is for a triplet ($ J=1 $),
figure 2b for a multiplet with $ J = 3 $ (seven components),
and figure 2c for a multiplet with $ J = 5 $ (eleven components).
For large $ J $,
$ S^{\prime} $ increases faster that $ T $,
and $ U^{\prime} $ increases faster than $ S^{\prime} $.

\vspace{5mm}

Figure 3: Plot of $ S $ (full line),
$ T $ (dotted line),
$ U $ (small-dashes line),
$ V $ (medium-dashes line),
$ W $ ( large-dashes line) and $ X $ (dashed-dotted line) against $ m_2 $,
with $ m_1 = 150 $ GeV kept fixed.
All figures are for a scalar multiplet with hypercharge $ -1/2 $.
Figure 3a is for a triplet,
figure 3b for a multiplet with $ J = 3 $,
and figure 3c for a multiplet with $ J = 5 $.
For large $ J $,
$ V $ and $ U $ become the largest parameters.
A partial cancellation occurs between $ S^{\prime} $
and $ S^{\prime \prime} $ for small $ m_2 $,
preventing $ S $ from becoming very large too.

\begin{thebibliography}{99}
%
\bibitem{T}
M.\ B.\ Einhorn, D.\ R.\ T.\ Jones and M.\ Veltman,
Nucl.\ Phys.\ {\bf B191}, 146 (1981).
%
\bibitem{S}
M.\ E.\ Peskin and T.\ Takeuchi,
Phys.\ Rev.\ Lett.\ {\bf 65}, 964 (1990);
Phys.\ Rev.\ D {\bf 46}, 381 (1992).
%
\bibitem{STU}
D.\ C.\ Kennedy and B.\ W.\ Lynn,
Nucl.\ Phys.\ {\bf B322}, 1 (1989);
W.\ J.\ Marciano and J.\ L.\ Rosner,
Phys.\ Rev.\ Lett.\ {\bf 65}, 2963 (1990);
D.\ C.\ Kennedy and P.\ Langacker,
Phys.\ Rev.\ Lett.\ {\bf 65}, 2967 (1990); {\bf 66}, 395(E);
Phys.\ Rev.\ D {\bf 44}, 1591 (1991);
G.\ Altarelli and R.\ Barbieri,
Phys.\ Lett.\ B {\bf 253}, 161 (1991).
%
\bibitem{beyondSTU}
I.\ Maksymyk, C.\ P.\ Burgess and D.\ London,
McGill Report No.\ McGill-93/13, 1993 (unpublished).
%
\bibitem{grinstein}
B.\ Grinstein and M.\ B.\ Wise,
Phys.\ Lett.\ B {\bf 265}, 326 (1991).
%
\bibitem{fit}
C.\ P.\ Burgess, S.\ Godfrey, H.\ K\"onig, D.\ London and I.\ Maksymyk,
McGill Report No.\ McGill-93/24, 1993 (unpublished).
%
\bibitem{li}
L.-F.\ Li,
Carnegie-Mellon Report No. CMU-HEP92-18, 1992 (unpublished).
%
\bibitem{decoupling}
T.\ Appelquist and J.\ Carrazone,
Phys.\ Rev.\ D {\bf 11}, 2856 (1975).
%
\end{thebibliography}
\end{document}